\documentclass[a4paper,11pt]{article}
\usepackage{pos}
\usepackage{amsmath}
\usepackage{dsfont}
\usepackage{bbm}
\usepackage{cleveref}
\usepackage{subcaption}
\usepackage[utf8]{inputenc}
\usepackage{mathrsfs}
\usepackage{amsfonts}

\DeclareMathOperator{\str}{str}

\newcommand{\upq}{\mathrm{u}}
\newcommand{\downq}{\mathrm{d}}
\newcommand{\strangeq}{\mathrm{s}}
\newcommand{\charmq}{\mathrm{c}}

\newcommand{\lightq}{\mathrm{l}}

\newcommand{\MeV}{\text{MeV}}

\newcommand{\fm}{\text{fm}}

\newcommand{\ISO}{\ensuremath{\mathrm{iso}}}
\newcommand{\SIB}{\ensuremath{\mathrm{sib}}}
\newcommand{\QED}{\ensuremath{\mathrm{qed}}}
\newcommand{\QCD}{\ensuremath{\mathrm{qcd}}}
\newcommand{\PHY}{\ensuremath{\mathrm{phys}}}

\title{Checks on QED and strong-isospin breaking corrections to $a_{\mu}^{\mathrm{HVP}}$}
\ShortTitle{Checks on QED and strong-isospin breaking corrections to $a_{\mu}^{\mathrm{HVP}}$}

\author*[a]{Andreas Risch}
\onbehalf{for the Budapest-Marseille-Wuppertal collaboration}

\affiliation[a]{Department of Physics, University of Wuppertal, Gaussstr. 20, 42119 Wuppertal, Germany}

\emailAdd{andreas.risch@uni-wuppertal.de}

\abstract{At the current levels of precision reached in the measurement of the muon g-2 by Fermilab, it is essential to control QED and strong isospin breaking corrections to the HVP contribution to the muon g-2. Here we present a number of cross-checks performed on the results for those corrections presented in the BMW’s collaboration 2020 computation.}

\FullConference{The 41st International Symposium on Lattice Field Theory (LATTICE2024)\\
 28 July - 3 August 2024\\
Liverpool, UK\\}

\begin{document}
\maketitle

\section{Introduction}
\label{sec:introduction}

At the current levels of precision reached in the measurement of the muon $g-2$ at Fermilab~\cite{Muong-2:2023cdq} it is essential to control QED and strong isospin breaking (SIB) corrections to the hadronic vacuum polarisation (HVP) contribution to the anomalous magnetic moment of the muon $a_{\mu}^{\mathrm{HVP}}$. The QED and SIB uncertainties obtained in~\cite{Borsanyi:2020mff} are already sufficiently small to reach the precision desired in the most recent publication~\cite{Boccaletti:2024guq}. However, we have performed a variety of cross-checks that confirm those earlier findings. We organise this work as follows: We summarise the lattice setup and comment on improvements with regard to the previous computation~\cite{Borsanyi:2020mff}. We present an autocorrelation analysis and a discussion of taste violations. We recap the definition of the isospin breaking decomposition of our work~\cite{Boccaletti:2024guq} and present an analysis for the kaon mass decomposition comparing three different separation schemes. Finally, we present a variety of cross-checks of the computation of the isospin breaking contributions to $a_{\mu}^{\mathrm{HVP}}$ published in~\cite{Boccaletti:2024guq}. A related discussion of the reconstruction of the long-distance Euclidean current-current correlation function by means of a lattice QCD computation is presented in~\cite{Frech:2004}.
 
\section{Lattice setup}
\label{sec:latticesetup}

\begin{figure}[h]
\centering
\begin{minipage}{0.64\textwidth}
\centering
\includegraphics[width=\textwidth]{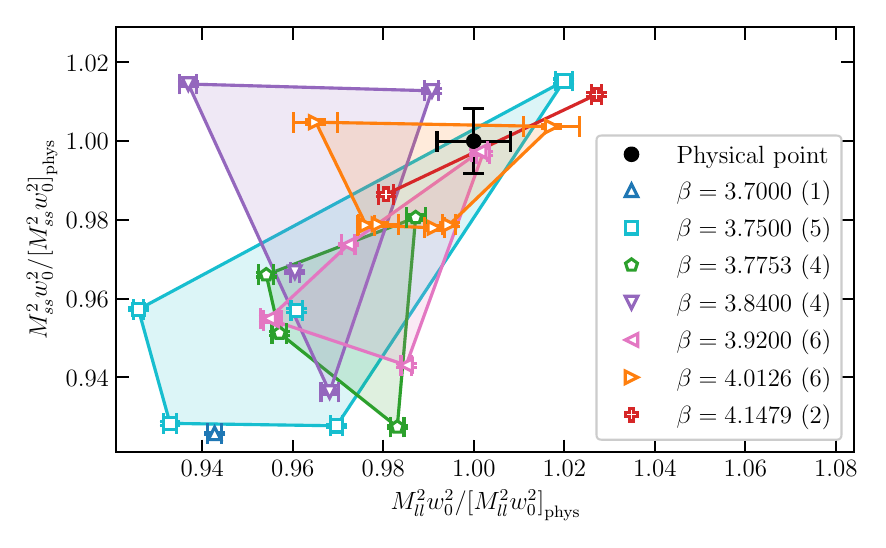}
\end{minipage}
\begin{minipage}{0.35\textwidth}\centering
\begin{tabular}{|c|c|c|c|c|}
\hline
$\beta$&$a$ [fm]&$L/a \times T/a $  \\
\hline
3.7000&0.1315&$48\times 64$\\
\hline
3.7500&0.1191&$56\times 96$\\
\hline
3.7753&0.1116&$56\times 84$\\
\hline
3.8400&0.0952&$64\times 96$\\
\hline
3.9200&0.0787&$80\times 128$\\
\hline
4.0126&0.0640&$96\times 144$\\
\hline
4.1479&0.0483&$128\times 192$\\
\hline
\end{tabular}
\end{minipage}
\caption{Landscape of ensembles. Left: The horizontal and vertical axes are the squared pseudo-scalar masses $M_{\lightq\lightq}^2$ and $M_{\strangeq\strangeq}^2$, which serve as proxies for the quark mass $m_{\lightq}$ and $m_{\strangeq}$, in units of the $w_{0}$-scale normalised with the corresponding physical values. The isospin symmetric physical point is shown in black. The uncertainties of this point result from the determination of $M_{\lightq\lightq}$, $M_{\strangeq\strangeq}$ and $w_{0}$ in physical units. Different colours denote different lattice spacings. Right: Inverse strong couplings $\beta$, lattice spacings $a$ and volumes of the different ensembles.\label{fig:landscape}}
\end{figure}

The computations are performed based on the tree-level improved Symanzik gauge action~\cite{Luscher:1984xn} in combination with a staggered fermion action. The gauge links in the staggered Dirac operator are stout smeared~\cite{Morningstar:2003gk} with $4$ smearing steps and a stout smearing parameter of $\rho = 0.125$. The Monte Carlo simulations for isospin symmetric QCD (QCD$_{\mathrm{iso}}$) possess $2+1+1$ dynamical quark flavours with degenerate light quark masses $m_{\upq}=m_{\downq}=m_{\lightq}$. In total, QCD$_{\mathrm{iso}}$ ensembles with $7$ different lattice spacings ranging from $0.1315\,\fm$ to $0.0483\,\fm$ are available as listed in \cref{fig:landscape}. Compared to the previous work on the anomalous magnetic moment of the muon~\cite{Borsanyi:2020mff} two ensembles with a lattice spacing of $a=0.0483\,\fm$ ($\beta=4.1479$) were added. The scale setting is based on the Wilson-flow observable $w_0$~\cite{BMW:2012hcm}. \Cref{fig:landscape} depicts also the landscape of ensembles in terms of the normalised pseudo-scalar masses $M_{\lightq\lightq}^2$ and $M_{\strangeq\strangeq}^2$ in units of the $w_{0}$-scale. $M^{2}_{qq}$ denotes the mass of the quark-connected pseudo-scalar meson with flavour $\bar{q}q$, which is a proxy for the quark mass $m_{q}$. The mass parameters $m_{\lightq}$ and $m_{\strangeq}$ of the light and strange quarks are scattered such that the physical point is enclosed by ensembles of the same lattice spacing. The charm mass parameter $m_{\charmq}$ is fixed by the ratio $m_{\charmq}/m_{\strangeq} = 11.85$, which was taken from the charm meson analysis in~\cite{McNeile:2010ji}. This value is within one percent of the most recent lattice average from FLAG~\cite{FlavourLatticeAveragingGroupFLAG:2021npn, FermilabLattice:2018est,EuropeanTwistedMass:2014osg,Chakraborty:2014aca}. More details on the physical values of $M_{\lightq\lightq}^2$, $M_{\strangeq\strangeq}^2$ and $w_{0}$, which are based on the experimental values of the masses of the $\pi^0$, $K^+$, $K^0$ and $\Omega^-$ hadrons, are discussed in \cref{sec:physicalpoint}.

\section{Autocorrelation analysis and taste violations}
\label{sec:autocorrelation}

\begin{figure}[h]
\centering
\includegraphics[width=0.8\textwidth]{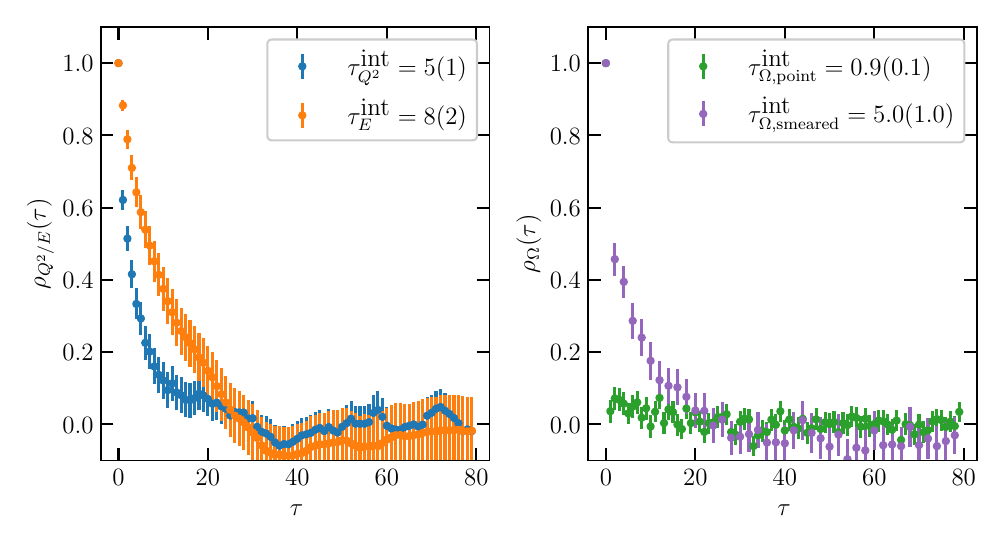}
\caption{Normalised autocorrelation functions $\rho(\tau)$ and integrated autocorrelation times $\tau^{\mathrm{int}}$ in units of configurations on one of  the two ensembles with finest lattice spacing ($\beta=4.1479$, $a=0.0483\,\fm$). Left: Energy density $E(w_0^2)$ and the squared topological charge $Q^{2}(w_0^2)$ evaluated at a gradient flow time of $t_{\mathrm{fl}}=w_0^2$. Right: $\Omega_\mathrm{VI}$ correlation function with point and smeared sources evaluated at $t= 1.5~\mathrm{fm}$.}
\label{fig:ACT}
\end{figure}

\Cref{fig:ACT} depicts normalised autocorrelation functions $\rho(\tau)$ and integrated autocorrelation times $\tau^{\mathrm{int}}$ in units of configurations for the smeared squared topological charge $Q^{2}(w_0^2)$ and the smeared energy density $E(w_0^2)$ as well as for the $\Omega$-baryon two-point function on one of the finest ensembles ($\beta=4.1479$, $a=0.0483\,\fm$). Using a parity preserving HMC algorithm simulating QCD the autocorrelation modes of the update algorithm can be classified according to parity~\cite{Schaefer:2010hu}. We study the parity even squared topological charge $Q^{2}$ instead of the parity odd $Q$ as the observables of interest are all parity even. In QCD parity odd expectation values vanish. The squared topological charge is computed based on the standard clover discretisation~\cite{Sheikholeslami:1985ij}. $Q^{2}(w_0^2)$ and $E(w_0^2)$ are evaluated at a gradient flow time $t_{\mathrm{fl}}=w_0^2$, which corresponds to a smearing radius of about $0.5\,\fm$. Autocorrelations of the $\Omega$-baryon correlation function based on the $\Omega_\mathrm{VI}$ discretisation, c.f.~\cite{Boccaletti:2024guq}, are studied for point and smeared sources. The correlation function is evaluated at $t\approx 1.5~\mathrm{fm}$, which is in the plateau region of the corresponding effective mass. The normalised autocorrelation functions and the integrated autocorrelation times were computed using the {\tt pyerrors} package~\cite{Joswig:2022qfe}. The integrated autocorrelation time of the smeared squared topological charge and the smeared energy density in units of configurations read $\tau_{Q^{2}}^{\mathrm{int}}=5(1)$ and $\tau_{E}^{\mathrm{int}}=8(2)$, whereas we find integrated autocorrelation times for the $\Omega$-baryon correlation function that are of similar and smaller magnitude. In the analysis, delete-one jackknife resampling is used to estimate statistical errors. To suppress autocorrelations between data from subsequent configurations a blocking procedure based on 48 blocks is used. For the finest ensembles ($\beta=4.1479$, $a=0.0483\,\fm$) this results in a block length of $B \approx 10\cdot \tau^{\mathrm{int}}_{\mathrm{Q}^{2}}$, i.e. the block length is sufficiently large compared to the relevant integrated autocorrelation times.

\begin{figure}[h]
\centering
\includegraphics[width=0.7\textwidth]{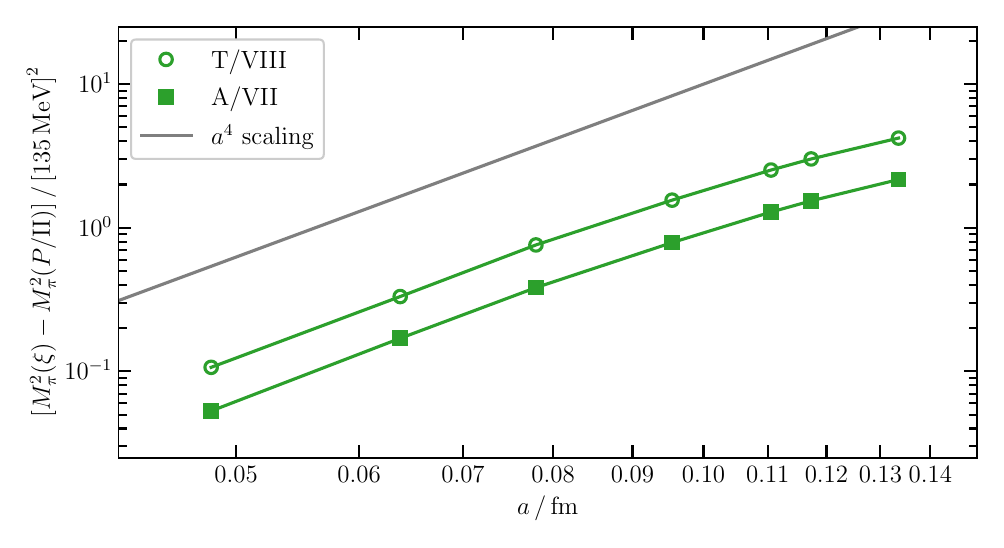}
\caption{Taste violations as a function of the lattice spacing $a$ for axial-vector $A$ and tensor $T$ tastes.\label{fig:taste_violation}}
\end{figure}

In the staggered fermion discretisation lattice artefacts are related to the taste symmetry violation of staggered fermions. As a result, pseudo-scalar mesons on the lattice become heavier than in the continuum and their masses depend on their taste quantum numbers. The masses of the pions $M^2_{\pi}(\xi)= M^2_{\lightq\lightq} + \Delta_{\mathrm{KS}}(\xi)$ are computed at the average valence quark mass $\tfrac{1}{3}(2m_{\lightq}+m_ {\strangeq})$. $\xi\in\{\mathrm{P},\mathrm{A},\mathrm{T},\mathrm{V},\mathrm{I}\}$ denotes one of the sixteen meson tastes of the $SU(4)$ taste group. $M^2_{\lightq\lightq}=M^2_{\pi}(\mathrm{P})$ is the squared mass of the pseudo-Goldstone pion, i.e. $\Delta_{\mathrm{KS}}(\mathrm{P})=0$. In \cref{fig:taste_violation} $\Delta_{\mathrm{KS}}$ is displayed for the $\mathrm{A}$ and $\mathrm{T}$ tastes as a function of the lattice spacing. We find that the taste violations decrease with approximately $a^{4}$ for small lattice spacings, which is also consistent with $\alpha_{\mathrm{s}}^{3} a^2$ scaling. $\alpha_{\mathrm{s}}$ is the strong coupling constant at the lattice cut-off scale. At the smallest lattice spacing ($\beta=4.1479$, $a=0.0483\,\fm$) $M_{\pi}(\mathrm{T})\approx 142\,\MeV$.  In~\cite{Boccaletti:2024guq} $\Delta_{\mathrm{KS}}(\mathrm{A})$ is used in the continuum extrapolation procedure to parametrise lattice artefacts.

\section{Physical point and isospin breaking decomposition}
\label{sec:physicalpoint}

In~\cite{Borsanyi:2020mff} the physical point of QCD+QED is defined by the location in parameter space where $M_{\pi^{0}}$, $M_{K^{0}}$, $M_{K^{+}}$, $M_{\Omega^{-}}$ and $\alpha$ adopt their physical values. The physical point can alternatively be translated into a scheme based on $\hat{M}^{2}=\frac{1}{2}(M^{2}_{\mathrm{uu}}+M^{2}_{\mathrm{dd}})$, $\Delta M^{2}=M^{2}_{\mathrm{dd}}-M^{2}_{\mathrm{uu}}$, $M^{2}_{\mathrm{ss}}$, $w_{0}$ and $\alpha$. The latter scheme is motivated by a computation in leading-order partially-quenched chiral perturbation theory coupled to photons~\cite{Bijnens:2006mk}, according to which $\hat{M}^{2}=\frac{1}{2}(M_{\mathrm{uu}}^{2}+M_{\mathrm{dd}}^{2})=M_{\pi^{0}}^{2}+\mathrm{NLO}$. $\Delta M^{2}=2B_{2}(m_{\downq}-m_{\upq})+\mathrm{NLO}$ is a measure for strong isospin breaking. We define the physical value of $\hat{M}^{2}$ by neglecting next-to-leading order effects such that $\hat{M}^{2}=M_{\pi^{0}}^{2}$, i.e.
\begin{align*}
[\hat{M}]_\PHY = 134.9768(5)~\mathrm{MeV}.
\end{align*}
The physical values of $\Delta M^2$ and $M_{ss}$ were computed in~\cite{Borsanyi:2020mff},
\begin{align*}
[\Delta M^2]_\PHY &=  13170(320)(270)[420]~\mathrm{MeV}^2 \\
[M_{ss}]_\PHY &= 689.89(28)(40)[49]~\mathrm{MeV},
\end{align*}
where the first error is statistical, the second is systematic and the third is the total. Compared to the previous work~\cite{Borsanyi:2020mff} the computation of the gradient flow scale $w_{0}$ was updated taking into account the additional ensembles at $0.048\,\fm$:
\begin{align*}
[w_0]_\PHY &=  0.17245(22)(46)[51]\,\fm.
\end{align*}
For more details on this update we refer to~\cite{Boccaletti:2024guq}.
%\cite{Wang:2024,Boccaletti:2024guq}

The scheme based on $\hat{M}^{2}$, $\Delta M^{2}$, $M^{2}_{\mathrm{ss}}$, $w_{0}$ and $\alpha$ can be used to decompose observables into isospin symmetric and isospin breaking contributions. In dimensionless units a point in parameter space is parametrised by $(\hat{M}w_0,M_{ss}w_0,\Delta M^2w_0^2,e)$. Then the physical points of QCD+QED, QCD and QCD$_{\mathrm{iso}}$ are defined by
\begin{align}
\begin{split}
\mathrm{QCD+QED}:\quad & ([\hat{M}w_0]_{\PHY},[M_{ss}w_0]_{\PHY}, [\Delta M^2 w_0^2]_\PHY, \alpha_\PHY) \\
\mathrm{QCD}:\quad & ([\hat{M}w_0]_{\PHY},[M_{ss}w_0]_{\PHY}, [\Delta M^2 w_0^2]_\PHY, 0) \\
\mathrm{QCD}_{\mathrm{iso}}:\quad & ([\hat{M}w_0]_{\PHY},[M_{ss}w_0]_{\PHY}, 0, 0).
\end{split}
\label{eq:scheme}
\end{align}
Observables can now be understood as functions of $(\hat{M}w_0,M_{ss}w_0,\Delta M^2w_0^2,e)$. The separation scheme allows us to decompose an observable computed in the full theory QCD+QED into $[O]_\PHY= [O]_\ISO + [O]_\SIB + [O]_\QED$, where $ [O]_\ISO $ is the isospin symmetric contribution evaluated in QCD$_{\mathrm{iso}}$, $[O]_{\SIB}=[O]_\QCD- [O]_\ISO$ is the strong isospin breaking contribution, where $[O]_\QCD$ is the QCD contribution, and $[O]_\QED= [O]_\PHY - [O]_\QCD$ is the electromagnetic contribution.

\begin{figure}[h]
\centering
\includegraphics[width=0.7\textwidth]{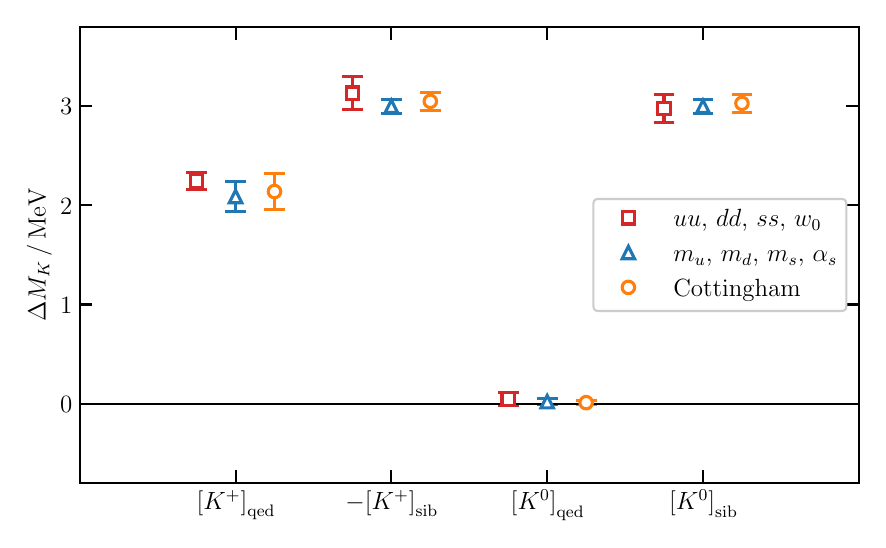}
\caption{Decomposition of the neutral and charged kaon masses in three different schemes: $(\hat{M}^{2},\Delta M^{2},M^{2}_{\mathrm{ss}},w_{0},\alpha)$ scheme, GRS scheme and Cottingham formula based scheme.\label{fig:kaondecomposition}}
\end{figure}

There is no unique scheme to separate the isospin symmetric contribution from isospin breaking contributions as QCD and QCD$_{\mathrm{iso}}$ are not realised in nature. Therefore, separation scheme ambiguities arise. To study these ambiguities we compare the kaon mass decomposition in three different schemes, in particular the scheme given above, the Gasser-Rusetsky-Scimemi (GRS) scheme~\cite{Gasser:2003hk,DiCarlo:2019thl} and a Cottingham-formula based decomposition~\cite{Stamen:2022uqh}. For the kaon mass decomposition in the scheme of \cref{eq:scheme} we find the following values:
\begin{align*}
[M_{K^{0/+}}]_\ISO = 494.55(31)~\mathrm{MeV} 
\quad\quad
\begin{aligned}
[M_{K^0}]_\SIB = +2.98(14)~\mathrm{MeV} \quad\quad [M_{K^0}]_\QED = 0.05(7)~\mathrm{MeV} \\
[M_{K^+}]_\SIB = -3.13(17)~\mathrm{MeV} \quad\quad [M_{K^+}]_\QED = 2.25(8)~\mathrm{MeV}.
\end{aligned}
\end{align*}
The results of the comparison are displayed in \cref{fig:kaondecomposition}, where values were taken from~\cite{Stamen:2022uqh,Giusti:2017dmp}. Overall, we find a good agreement of the different schemes, which indicates an equivalence of schemes at the given level of precision. We note that our isospin symmetric kaon mass agrees with the GRS scheme value $[M_{K^{0/+}}]_{\ISO,\mathrm{GRS}} = 494.6(1)\,\MeV$~\cite{DiCarlo:2019thl}.  

\section{Verification of isospin breaking contributions}

In the most recent work~\cite{Boccaletti:2024guq} only the isospin symmetric contribution to $a_{\mu}^\mathrm{light}$ and $a_{\mu}^\mathrm{disc}$ were updated, whereas the isospin breaking contributions were taken from the previous work~\cite{Borsanyi:2020mff}. Nevertheless, in order to validate the previous procedures several additional analyses and improvements were performed.

\begin{figure}
\centering
\includegraphics[width=0.5\textwidth]{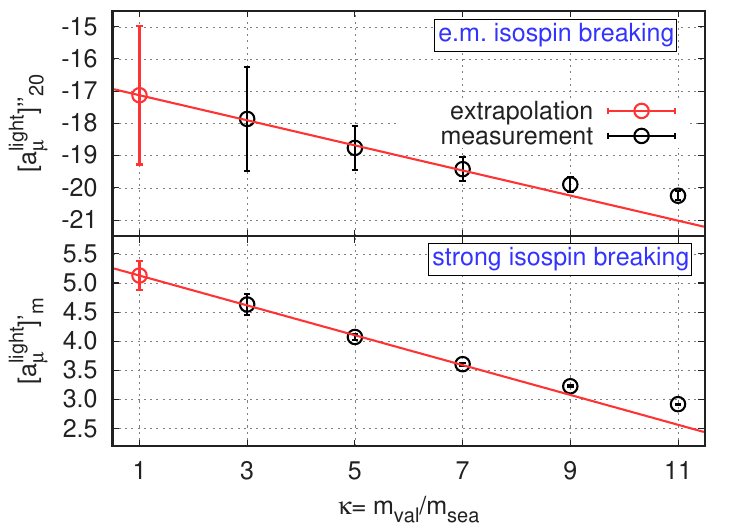}
\caption{Extrapolation procedure for $[a_{\mu}^\mathrm{light}]'_m$ and $[a_{\mu}^\mathrm{light}]''_{20}$ for $\beta=3.7000$ in the previous work~\citep{Borsanyi:2020mff}.\label{fig:ibjj}}
\end{figure}
\begin{figure}
\centering
\begin{minipage}{0.49\textwidth}
\centering
\includegraphics[width=\textwidth]{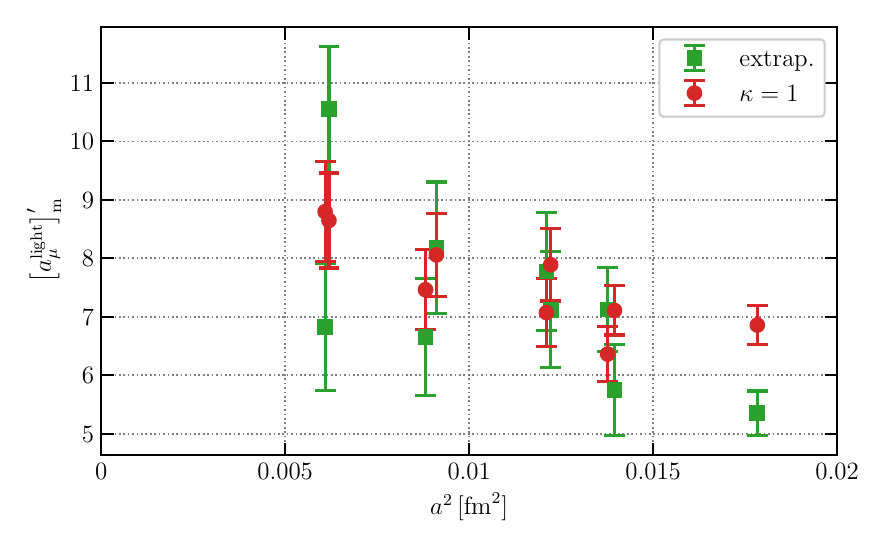}
\caption{Strong-isospin breaking contribution to $a_{\mu}^{\text{light}}$. Comparison of computations of $[a_{\mu}^\mathrm{light}]'_m$ based on a chiral extrapolation and based on low-mode averaging applied at $\kappa=1$.\label{fig:l_sib_xchk}
}
\end{minipage}
\hfill
\begin{minipage}{0.49\textwidth}
\centering
\includegraphics[width=\textwidth]{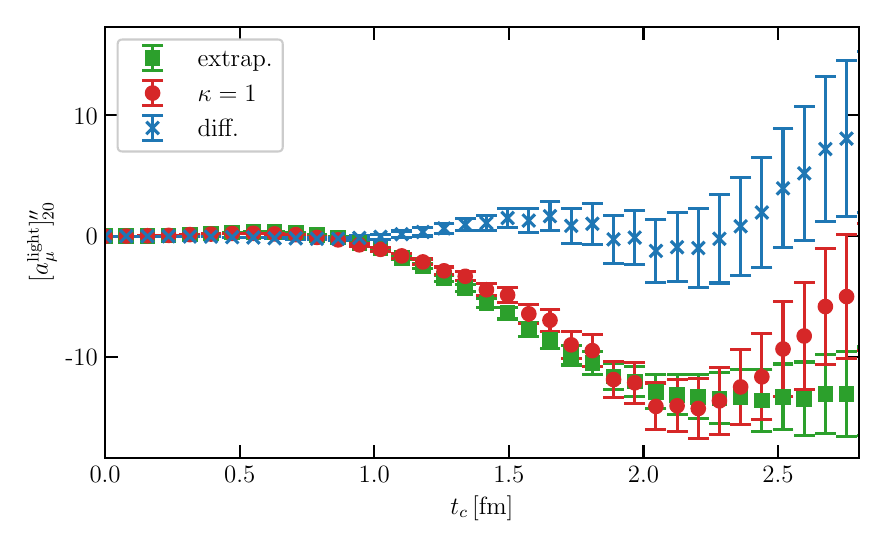}
\caption{$[a_{\mu}^\mathrm{light}]''_{20}$ computed on a single configuration at $a= 0.0787\,\fm$ as a function of the upper limit of integration $t_c$. Comparison of computations based on a chiral extrapolation and based on low-mode averaging applied at $\kappa=1$.\label{fig:l_q20_xchk}
}
\end{minipage}
\end{figure}

The strong isospin breaking (SIB) and the valence QED contributions to $a_{\mu}^\mathrm{light}$ are described by the derivatives
\begin{align*}
[a_{\mu}^\mathrm{light}]'_m &\equiv  m_l \, \frac{\partial
    [a_{\mu}^\mathrm{light}]}{\partial\, \delta
    m} \Big|_{\delta m=0} &     [a_{\mu}^\mathrm{light}]''_{20} &\equiv \frac{1}{2} \frac{\partial^2
    [a_{\mu}^\mathrm{light}]}{\partial e_v^2}  \Big|_{e_v=0},
\end{align*}
where $\delta m = m_{\downq} - m_{\upq}$ and $m_{\lightq} = \frac{1}{2}( m_{\upq} + m_{\downq})$. $e_v$ denotes the charge of the observable's valence quarks. In the previous setup~\cite{Borsanyi:2020mff} these computations were based on chiral extrapolations as depicted in~\cref{fig:ibjj}. Measurements were performed at valence quark masses $\kappa \cdot m_{l}$ with $\kappa={3,5,7,9,11}$, where $m_{l}$ is fixed to the sea quark mass, and a linear chiral extrapolation to $\kappa=1$ based on $\kappa={3,5,7}$ was performed gauge configuration by gauge configuration.

In the current setup~\cite{Boccaletti:2024guq} the discrete mass derivative and the chiral extrapolation in the computation of $[a_{\mu}^\mathrm{light}]'_m$ were replaced by an exact mass derivative in combination with a direct computation at $\kappa=1$ based on low-mode averaging~\cite{Giusti:2004yp,DeGrand:2004qw}. The latter technique was already used to reduce the noise in the computation of the isospin symmetric contribution. A comparison is show in \cref{fig:l_sib_xchk} which confirms the results of the previous publication~\cite{Borsanyi:2020mff}.

For the valence QED contribution $[a_{\mu}^\mathrm{light}]''_{20}$ the chiral extrapolation was cross-checked on a single gauge configuration from one of the $a= 0.0787\,\fm$ ensembles based on a low-mode averaging computation. \Cref{fig:l_q20_xchk} displays $[a_{\mu}^\mathrm{light}]''_{20}$ as a function of the time cut $t_c$, which is the upper limit of the integration of the Euclidean time current-current correlation function, c.f.~\cite{Boccaletti:2024guq}, showing a difference of the two methods that is compatible with zero.

\begin{figure}[h]
\centering
\includegraphics[width=0.5\textwidth]{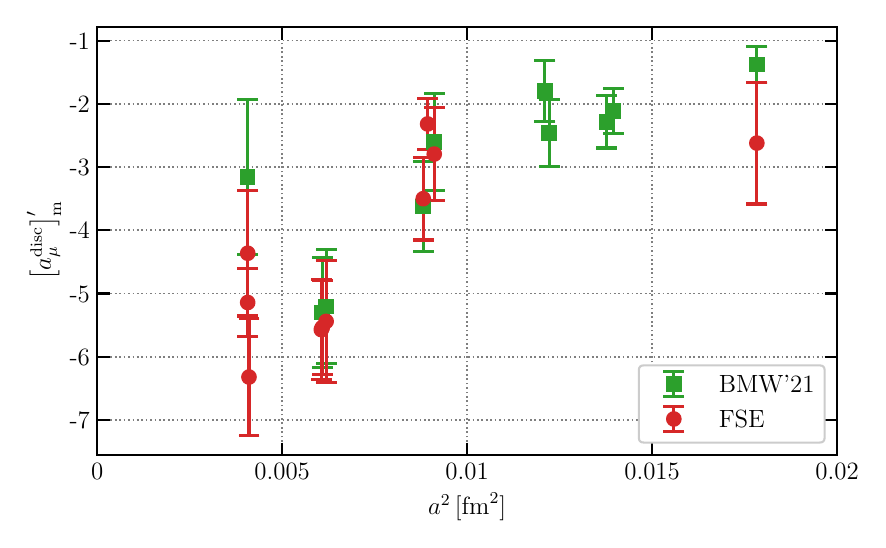}
\caption{Comparison between the standard stochastic estimator and frequency-splitting estimator of the strong isospin breaking contribution $[a_{\mu}^\mathrm{disc}]'_{m}$. \label{fig:di_sib_xchk}
}
\end{figure}

In the previous setup~\cite{Borsanyi:2020mff} the SIB contribution to
$a_{\mu}^\mathrm{disc}$ was computed performing a discrete derivative with respect to $\delta m$ based on $1.0\cdot m_l$ and $0.9\cdot m_l$. This setup was replaced by a computation based on a frequency-splitting estimator~\cite{Giusti:2019kff} in combination with an exact mass derivative. A comparison between the two methods is shown in \cref{fig:di_sib_xchk}. The number of required random source vectors for the evaluation of the stochastic estimator decreased by one to two orders of magnitude such that we obtained results with comparable error sizes at reduced cost.

\section{Conclusion and Outlook}

At the current level of precision reached in the measurement of the anomalous magnetic moment of the muon at Fermilab~\cite{Muong-2:2023cdq} it is essential to control QED and strong isospin breaking corrections to the HVP contribution to the muon g-2. In this work we discussed cross-checks and improvements presented in~\cite{Boccaletti:2024guq} that were performed and introduced compared to the previous publication~\cite{Borsanyi:2020mff}. In particular, two new ensembles with lattice spacing $a=0.0483\,\mathrm{fm}$ were added to the analysis. We performed an autocorrelation analysis and presented the results for the finest ensembles showing that effects due to autocorrelations are under control. We further studied the taste symmetry breaking causing lattice artefacts in the staggered fermion discretisation. We also computed the isospin breaking decomposition of $M_{K}$ and found a consistency between three separation schemes: Our scheme motivated by a leading-order partially-quenched chiral perturbation theory computation coupled to photons, the GRS scheme and a Cottingham formula based scheme. This consistency is an indication of the equivalence of the schemes at the current level of precision. We further improved the computation of leading isospin breaking corrections to $a_{\mu}^{\mathrm{HVP}}$ replacing computations based on chiral extrapolations by direct computations at the desired quark masses applying noise reduction techniques. We verified the results published in the previous publication~\cite{Borsanyi:2020mff} .

\acknowledgments
The authors gratefully acknowledge the Gauss Centre for Supercomputing (GCS) e.V., GENCI (grant 502275) and EuroHPC Joint Undertaking (grant EXT-2023E02-063) for providing computer time on the GCS supercomputers SuperMUC-NG at Leibniz Supercomputing Centre in München, HAWK at the High Performance Computing Center in Stuttgart and JUWELS and JURECA at Forschungszentrum Jülich, as well as on the GENCI supercomputers Joliot-Curie/Irène Rome at TGCC, Jean-Zay V100 at IDRIS, Adastra at CINES and on the Leonardo supercomputer hosted at CINECA. 

\bibliographystyle{JHEP}
\bibliography{proceedings.bib}

\end{document}